\documentclass[conference]{IEEEtran}
\IEEEoverridecommandlockouts
\usepackage{cite}
\usepackage{amsmath, amssymb, amsfonts}
\usepackage{algorithmic}
\usepackage{graphicx}
\usepackage{textcomp}
\usepackage{xcolor}
\def\BibTeX{{\rm B\kern-.05em{\sc i\kern-.025em b}\kern-.08em
    T\kern-.1667em\lower.7ex\hbox{E}\kern-.125emX}}
\begin{document}

\title{SQL-to-Schema Enhances Schema Linking in Text-to-SQL\\
}

\author{\IEEEauthorblockN{1\textsuperscript{th} Sun Yang}
\IEEEauthorblockA{\textit{Peking University} \\
Beijing, China \\
2201210484@stu.pku.edu.cn}
\and
\IEEEauthorblockN{2\textsuperscript{th} Qiong Su}
\IEEEauthorblockA{\textit{Guizhou University} \\
Guizhou, China \\
suqiaong.gzu@gmail.com}
\and
\IEEEauthorblockN{3\textsuperscript{th} Zhishuai Li}
\IEEEauthorblockA{\textit{SenseTime Research} \\
Shanghai, China \\
lizhishuai@sensetime.com}
\and
\IEEEauthorblockN{4\textsuperscript{th} Ziyue Li}
\IEEEauthorblockA{\textit{University of Cologne}\\
Cologne, Germany \\
zlibn@wiso.uni-koeln.de}
\and
\IEEEauthorblockN{5\textsuperscript{th} Hangyu Mao}
\IEEEauthorblockA{\textit{SenseTime Research} \\
Shanghai, China \\
maohangyu@sensetime.com}
\and
\and
\IEEEauthorblockN{6\textsuperscript{th} Chenxi Liu}
\IEEEauthorblockA{\textit{Nanyang Technological University} \\
Singapore \\
chenxi.liu@ntu.edu.sg}
\and
\IEEEauthorblockN{7\textsuperscript{th} Rui Zhao}
\IEEEauthorblockA{\textit{SenseTime Research} \\
Shanghai, China \\
zhaorui@sensetime.com}
}

\maketitle

\begin{abstract}
In sophisticated existing Text-to-SQL methods exhibit errors in various proportions, including schema-linking errors (incorrect columns, tables, or extra columns), join errors, nested errors, and group-by errors. Consequently, there is a critical need to filter out unnecessary tables and columns, directing the language model's attention to relevant tables and columns with 
 schema-linking, to reduce errors during SQL generation. Previous approaches have involved sorting tables and columns based on their relevance to the question, selecting the top-ranked ones for sorting, or directly identifying the necessary tables and columns for SQL generation. However, these methods face challenges such as lengthy model training times, high consumption of expensive GPT-4 tokens in few-shot prompts, or suboptimal performance in schema linking. Therefore, we propose an inventive schema linking method in two steps: Firstly, generate an initial SQL query by utilizing the complete database schema. Subsequently, extract tables and columns from the initial SQL query to create a concise schema. Using CodeLlama-34B, when comparing the schemas obtained by mainstream methods with ours for SQL generation, our schema performs optimally. Leveraging GPT-4, our SQL generation method achieved results that are comparable to mainstream Text-to-SQL methods on the Spider dataset.
\end{abstract}


\section{Introduction}
In the era of the Digital Revolution, where data drives human activities, querying complexities hinder access for non-experts. Existing interfaces cater to experts or offer limited capabilities. To democratize data access, breaking technical barriers is crucial. This brings us to Text-to-SQL, a system that translates natural language queries into the underlying database language, enabling broader data access and usability\cite{a1}.The cross-domain dataset for Text-to-SQL, Spider\cite{spider}, followed suit, attracting numerous researchers from the Natural Language Processing and Data Science communities. 

Moreover, most previous work assumes that
user queries contain exact column names and entries. However, it is unrealistic that users always formulate their questions with exact column names and string entries in the table. To tackle this issue, when scaleability and privacy are not of a concern, the system needs to search databases to better understand what the user is querying\cite{TypeSQL}. Therefore, schema linking emerged. Schema linking is a specialized form of entity linking that associates phrases in a given question with column or table names in the database schema.

SLSQL\cite{SLSQL}illustrates the crucial role of schema linking in enhancing SQL parsing performance. This process entails identifying the necessary tables and columns within a specific database schema and question context for generating effective SQL queries. Through relation-aware self-attention, RAT-SQL\cite{RAT-SQL} introduces a unified framework to tackle schema encoding and linking challenges, learning schema and question representations jointly based on their alignment and schema relations. SemQL\cite{SemQL}presents a neural approach for complex and cross-domain Text-to-SQL, aiming to address the lexical problem and the mismatch problem with schema linking and intermediate representation. With the advent of large language models, inspired by SLSQL, RAT-SQL, SemQL, many researchers are re-examining the role of schema linking in Text-to-SQL. Various fine-tuning\cite{RESDSQL} and prompting\cite{DIN-SQL, C3} Text-to-SQL methods with schema linking module appear spontaneously. However, these methods suffer from high training time costs and token consumption issues.

Historically, Text-to-SQL methods fell short in recognizing the significance of Text-to-SQL, either neglecting schema-linking or providing schema linking methods with limited impact on the overall Text-to-SQL task. Hence, we propose a novel approach: utilizing the complete schema and the question to compose a prompt for large language models to generate an initial SQL query, subsequently, parsing the initial SQL to extract columns and tables to form the linking schema. In summary, our contributions are: 
\begin{itemize}
\item The first to propose extracting the linking schema from the initial SQL, namely, SQL-to-Schema. Also, the first to define evaluation metrics for the schema linking module, enabling researchers to swiftly validate the effectiveness of schema linking without waiting for SQL generation.
\item Compared to linking schemas obtained through other methods, using the linking schema we extracted for SQL generation achieved the optimal execution accuracy when using codallma-34B.
\item  When our schema linking is combined with our complete Text-to-SQL approach, using GPT-4, it outperforms all zero shot and few shot prompts approaches, indicating that the benefits brought by our schema linking can propagate to the overall Text-to-SQL task.
\end{itemize}

\section{Related Work}

\subsection{Customized Machine Learning Fine-tuning Methods}
The traditional machine learning methods entail the concatenation of questions and database schemas into embeddings, training  models to obtain softmax probabilities of relevance between these questions and columns. The tables and columns with the highest probabilities are selected as linking schemas. For example, RESDSQL\cite{RESDSQL}selects the top 4 tables with the highest relevance, and each table includes the top 5 most relevant columns. These approaches encounter challenges such as a lack of high-quality training data, high training time cost, and diminished quality of linking schemas when applied to cross-domain datasets.
\subsection{Stimulating general LLM with prompting}
The prompting method is mainly divided into two types: Few-shot prompting, such as DIN-SQL\cite{DIN-SQL}, provides multiple schema linking examples within the prompt, it leverages the large model's in-context learning ability to enable the model to identify the precise schema for a given question. Zero-shot prompting, such as C3\cite{C3}, sorts the correlation between tables and columns from the database schema and the words in the question, and then selects the most relevant tables and columns as linking schema. Since few-shot prompting relies on examples tailored for the Spider dataset, it performs poorly on other cross-domain databases. Zero-shot prompting may result in low-quality schemas because some columns may lack correlation with the words in the question, yet they might still be necessary for SQL generation. Additionally, these current  prompting methods require the use of multiple complex modules with long prompts, which incur a significant amount of expensive GPT-4 token costs.

Below is an overview of currently representative Text-to-SQL or schema linking methods:\begin{itemize}
\item RESDSQL\cite{RESDSQL}:it improves Text-to-SQL parsing with a cross-encoder ranking schema items. Its framework decouples schema linking and skeleton parsing, demonstrating promising performance and robustness on Spider and variants.
\item DIN-SQL\cite{DIN-SQL}:it advances database interfaces, proposing effective LLM-based few-shot prompting for Text-to-SQL tasks. Achieves superior performance on Spider and BIRD benchmarks, addressing complexity and schema linking challenges.
\item C3\cite{C3}:it consists of three key components: Clear Prompting, Calibration with Hints, and
Consistent Output, which are corresponding to the model input, model bias and model
output respectively. It provides a systematic treatment for zero-shot Text-to-SQL.
\item Dail-SQL\cite{Dail-SQL}:it systematically explores prompt engineering, employs open-source LLMs for Text-to-SQL, compares prompt efficiency.
\item ChatGPT\cite{chatgpt}:it utilizes a fixed Text-to-SQL prompt for ChatGPT, emphasizing SQLite compatibility. The prompt supports both single-turn and multi-turn scenarios, requiring database and question information.
\item Graphix-T5\cite{Graphix-T5}:it explores two dimensions in Text-to-SQL parsers: incorporating relational structures into neural networks and leveraging pre-trained models. Introducing GRAPHIX-T5, the model embeds structural information via a GRAPHIX layer in the T5 encoder-decoder architecture.
\item T5-base + Tok\cite{T5-base + Tok}:it explores LM generalization in Text-to-SQL parsing using the realistic and challenging cross-database Spider benchmark. It introduces two easy-to-implement techniques at token and sequence levels, significantly improving LM generalization, especially for compositional tasks.
\item RATSQL+GAP+NatSQL\cite{RATSQL+GAP+NatSQL}:it proposes a simplified SQL intermediate representation, enhancing natural language inference, enabling executable SQL generation.
\end{itemize}

\section{Methodology}
Using Codellama-34B, compared to the SQL generated using the complete database schema without using schema linking module, the SQLs generated from the linking schemas extracted by the C3, DIN-SQL, and RESDSQL methods exhibits only a modest improvement in execution accuracy, with respective gains of 0.01, 0.001, and 0.008. The Dail-SQL\cite{Dail-SQL} method employs initial SQL filtering for few-shot examples, and achieves good performance on the Text-to-SQL task. Inspired by Dail-SQL, we seek to enhance Text-to-SQL in execution accuracy by harnessing the power of an initial SQL query, namely, SQL-to-Schema. Initially, we generate a preliminary SQL using the complete database schema and then employ SQL parsing methods to extract tables and columns, forming our linking schema. Figure 1 includes three modules: Initial SQL Generation, SQL Parse, and SQL Generation. It illustrates a complete example of SQL-to-Schema.

\begin{figure}[htbp]
\centerline{\includegraphics[width=1\columnwidth]{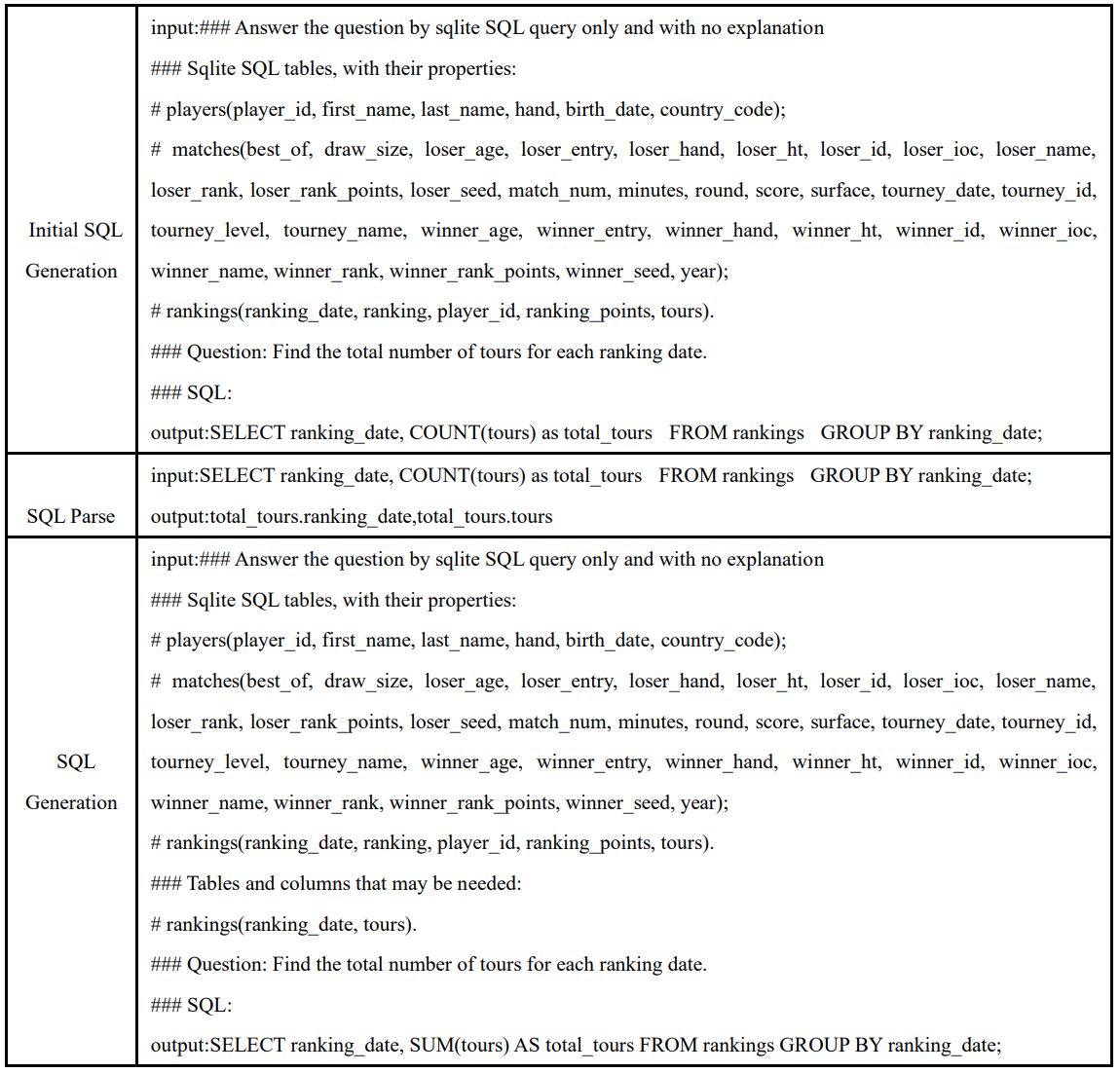}}
\caption{The complete example of SQL-to-Schema}
\label{fig}
\end{figure}

\subsection{Evaluation Metrics}
To assess the quality of linking schemas, we consider that if a schema does not include all the tables and columns required to generate the SQL, the correct SQL cannot be generated. Given that each question in the Spider dataset involves a maximum of 4 tables, and both the C3 and RESDSQL methods select the top 4 tables from their sorted schemas, along with the top 5 columns for each table, to form the linking schema. At the same time, DIN-SQL and our method extracts schemas containing no more than 4 tables. Therefore, we parse the gold SQL to extract the ground truth labels for tables and columns. We define table-recall@4, where the tables in the linking schema include all the tables involved in the gold SQL. In this case, the linking schema is considered a true schema; otherwise, it is considered a false schema. Since gold SQL and DIN-SQL schema includes some columns derived from aggregate functions, such as count(*), and the columns from C3 and RESDSQL methods are directly from the database schema without including count(*), the linking schemas obtained by C3 and RESDSQL may include all the columns needed to generate the SQL. Therefore, defining recall from the column perspective may affect a fair comparison. In the experimental section, we will use table-recall@4 as one of the evaluation metrics.

Due to the fact that each question in the Spider dev dataset provides only one gold SQL, while each database contains numerous foreign keys, a question may have multiple gold SQLs, each involving different tables. Therefore, evaluating the quality of schema linking using only table-recall@4 is insufficient. To address this, we introduce the SQL generation module, as depicted in Figure 1, to combine the schemas obtained from different methods with the SQL generation module. We compare the quality of different schema linking schemes using the generated SQL We employ the Text-to-SQL evaluation method mentioned in the Spider dataset, namely, execution accuracy\cite{EA}, and supplement it as an additional evaluation metric for subsequent experiments.

\subsection{Introduction to Each Module}
Initial SQL Generation. Numerous studies\cite{prompt, chatgpt, C3} have investigated prompting strategies for the Text-to-SQL task, conducting comprehensive comparisons of various prompt construction strategies for databases and demonstrations across zero-shot, single-domain, and cross-domain Text-to-SQL scenarios. Leveraging the insights from these prompting research efforts, we ultimately designed the Initial SQL Generation(ISG) prompt illustrated in Figure 1.

SQL Parse. This module is employed to extract the tables and columns labels from the gold SQL, simultaneously used to extract columns and tables from the SQLs of our other modules to form linking schemas. Due to the Spider dataset's gold SQL being annotated by multiple students and engineers at Yale University without a standardized SQL specification, using current SQL parsing packages might overlook columns. Therefore, we have designed the following SQL parse(SP) algorithm:
\begin{itemize}
\item Utilize sql metadata  package to parse the linking tables from the target SQL.
\item Split the target SQL into a target list by punctuation marks.
\item Iterate through all columns of the linking tables, and if a column matches an element in the target list, consider it as a linking column. 
\item If there is an asterisk (*) and the linking table does not match any linking column, consider all columns of this table as linking columns.
\end{itemize}

SQL Generation. In order to further leverage the power of SQL-to-Schema, we iteratively utilize the SQL generation and parsing modules multiple times to enhance the quality of linking schemas and final SQLs. Therefore, the input of SQL Generation(SG) module for this stage includes the linking schema output from the previous phase, the complete database schema, and the question. The output of this stage serves as the input for the SQL Parse stage in the next round. The advantages of this prompt format will be elaborated upon in the experimental section.

Self-Consistency Voting. Previous research \cite{voting}takes advantage of the fact that multiple prompts can be used to specify a single task, and propose to regularize prompt consistency, encouraging consistent predictions over this diverse set of prompts. Inspired by this, our Initial SQL Generation and SQL Generation employ distinct prompts. Consequently, we conduct a vote among all generated SQL queries for the same question, selecting the most consistent SQL as the final output for Text-to-SQL, a process referred to as self-consistency voting (SCV).

Our complete algorithm framework is illustrated in Figure 2. For clarity, we abbreviate our generated SQL and linking schema as follows:
\begin{itemize}
\item  SQL0: Obtained after ISL.
\item Schema1: Obtained after ISL, SP.
\item SQL1: Obtained after ISL, SP, SP, SG.
\item Schema2: Obtained after ISL, SP, SG, SP.
\item SQL2: Obtained after ISL, SP, SG, SP, SG.
\item Schema3: Obtained after ISL, SP, SG, SP, SG, SP.
\item SQL3: Obtained after ISL, SP, SG, SP, SG, SP, SG.
\item SCVSQL: Obtained after SCV using SQL0, SQL1, SQL2.
\end{itemize}

\begin{figure}[htbp]
\centerline{\includegraphics[width=0.9\columnwidth]{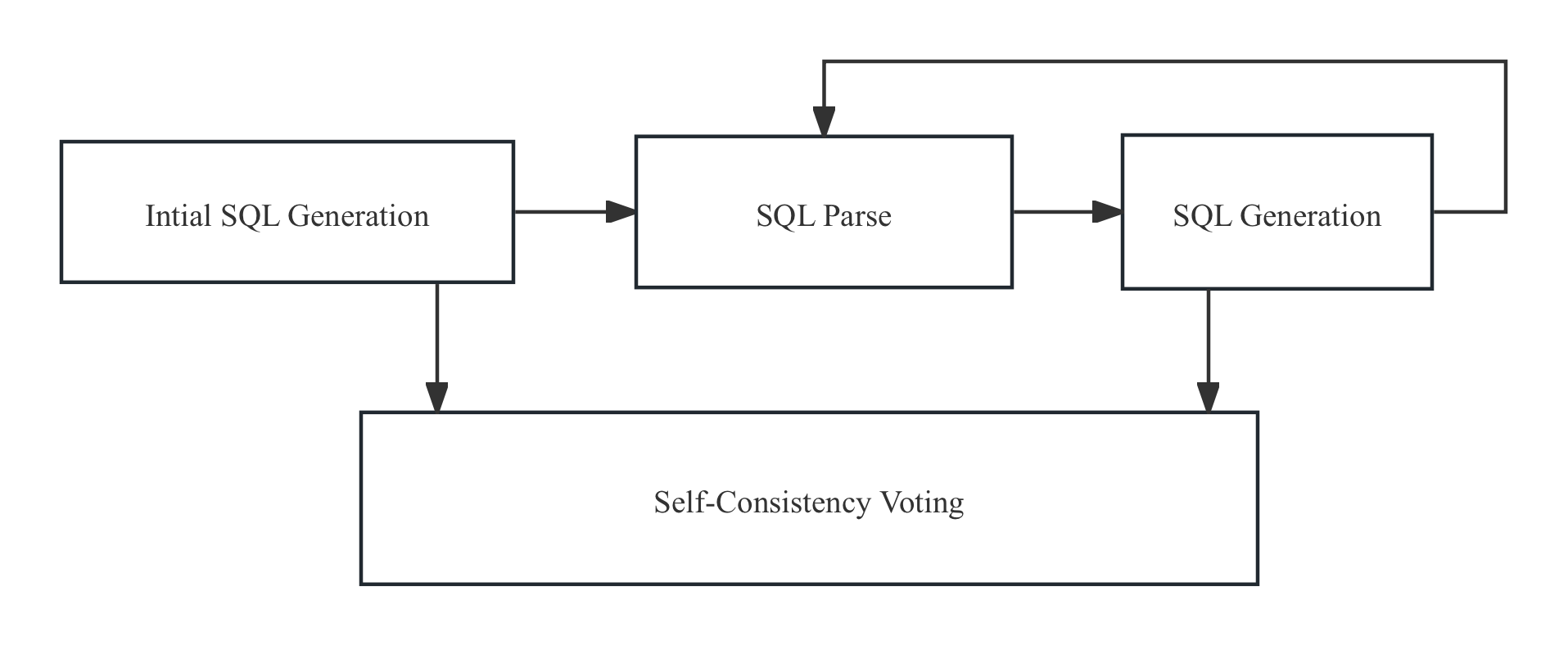}}
\caption{The complete schema linking and Text-to-SQL algorithm framework.}
\label{fig}
\end{figure}

\section{Experiments and Analysis}
\subsection{Experiment One}
Firstly, we compare the schemas obtained by the DIN-SQL, C3, and RESDSQL methods, as well as our generated linking schemas, using table-recall@4. The experimental results are presented in Table 1.

\begin{table}[htbp]
\caption{ The schema linking results of table-recall@4 on the Spider dev datasets}
\begin{center}
\begin{tabular}{|c|c|}
\hline 
Schema & table-recall@4 \\
\hline
DIN-SQL Schema & 0.88 \\
\hline
C3 Schema & 0.932 \\
\hline
RESDSQL Schema & 0.938 \\
\hline
Schema1 & 0.95 \\
\hline
Schema2 & 0.978 \\
\hline
Schema3 & 0.981 \\
\hline
\end{tabular}
\label{tab1}
\end{center}
\end{table}

When replicating the schema linking methods of DIN-SQL and C3 and implementing our method with Codellama-34B, and incorporating schemas obtained from the RESDSQL method for comparison using table-recall@4, Our method achieved state-of-the-art performance in this metric. Schema1, Schema2 and Schema3 significantly outperformed previous schema linking methods, setting a new benchmark. This implies the feasibility of SQL-to-Schema and inspires us to explore the use of intermediate step outputs to enhance the performance of large models in other NLP tasks.

\subsection{Experiment Two}
To further validate the feasibility of SQL-to-Schema, the obtained linking schemas from the aforementioned experiments were input into the SQL generation module depicted in Figure 1. SQLs were generated using Codellama-34B, and the execution accuracy of SQLs was employed as a measure to represent the quality of schema linking. The experimental results are presented in Table 2.

When directly using the prompt of Initial SQL Generation module combined with the linking schemas, some generated final SQLs contains errors, although the initial SQLs corresponding to them are correct. Through experimental analysis of the results, we observed that when all columns of a table in the linking schema appear in the SELECT clause of the gold SQL, the  generated  final SQL tends to use "SELECT *" which may output columns in that table that are unrelated to the question. Additionally, some linking schemas lack necessary tables and columns, resulting in errors in the generated SQL. Examples of these errors are illustrated in Figure 3.
\begin{figure}[htbp]
\centerline{\includegraphics[width=1\columnwidth]{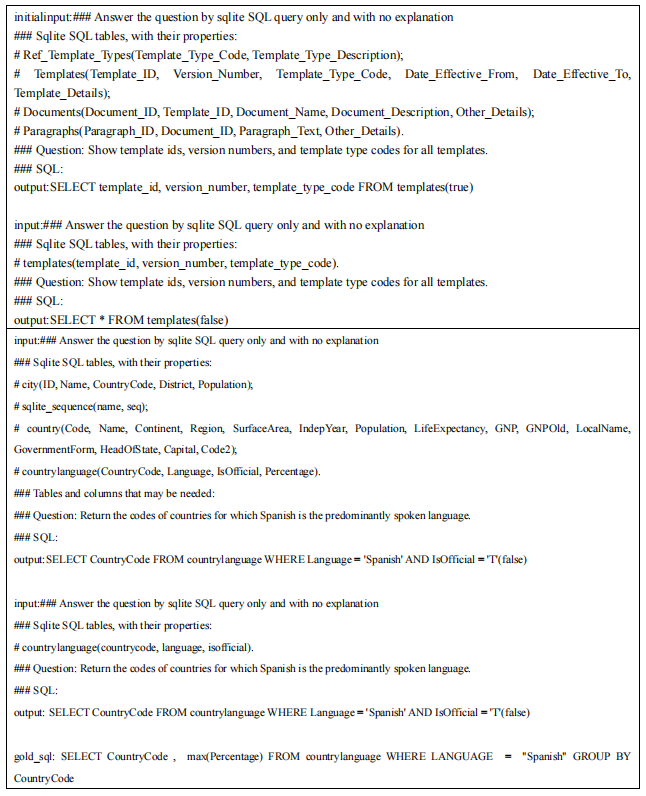}}
\caption{The single schema prompt error examples.}
\label{fig}
\end{figure}
Therefore, in subsequent SQL generation, we use the linking schema as a reference while including the complete database schema in the prompt, as illustrated in the SQL generation module in Figure 1.

\begin{table}[htbp]
\caption{ The results of SQL Execution Accuracy on the Spider dev dataset}
\begin{center}
\begin{tabular}{|c|c|}
\hline
Approaches&  EA\\
\hline
DIN-SQL schema +SQL Generation+codellama-34B& 0.723\\
\hline
C3 schema +SQL Generation+codellama-34B& 0.730\\
\hline
RESDSQL schema +SQL Generation+codellama-34B& 0.732\\
\hline
SQL0 +codellama-34B& 0.722\\
\hline
SQL1 +codellama-34B& 0.748\\
\hline
SQL2 +codellama-34B& 0.750\\
\hline
SQL3 +codellama-34B& 0.753\\
\hline
\end{tabular}
\label{tab1}
\end{center}
\end{table}

As shown in Table 2, SQL3 achieved the best performance on Codellama-34B, with a score of 0.753. SQL1, which only utilized the initial SQL, demonstrated a performance improvement of 0.026 compared to SQL0. SQL2 utilized SQL1 to extract linking schemas. SQL3 further utilized SQL2 to extract linking schemas. The execution accuracy continued to improve, indicating the effectiveness of the SQL-to-Schema strategy under two SQL generation prompts. Under the same experimental conditions except for the differing linking schemas, Table 2 demonstrates that the SQL-to-Schema strategy extracts higher-quality linking schemas.

\subsection{Experiment Three}

Since the schema linking module ultimately serves Text-to-SQL research task in both the NLP and DB communities, we cannot solely evaluate the quality of linking schemas and the feasibility of SQL-to-Schema from the perspective of the schema linking module. We must also consider whether the combination of our schema linking module and other functional modules related to Text-to-SQL tasks can bring benefits, namely, whether the gains from schema linking can be positively transferred to SQL generation. Therefore, we utilize GPT-4 and GPT-4 turbo to run all modules in Figure 2, generating SQLs, and compare them with the currently best-performing fine-tuning models and zero shot and few shots prompting  Text-to-SQL methods. While Schema1, Schema2, and Schema3 exhibit gradual improvement, considering that the table-recall@4 of Schema2 is approaching convergence, Schema3 would require more tokens and time. Therefore, for subsequent SQL generation, we utilized eight sets of experimental configurations:
\begin{itemize}
\item SQL0+GPT4.
\item SQL0+GPT4-turbo.
\item SQL1+GPT4.
\item SQL1+GPT4-turbo.
\item SQL2+GPT4.
\item SQL2+GPT4-turbo.
\item SCVSQL+GPT4.
\item SCVSQL+GPT4-turbo.
\end{itemize}
The omparison of Text-to-SQL experiments is shown in Table 3.

\begin{table}[htbp]
\caption{The Comparison of Execution Accuracy in Text-to-SQL Methods}
\begin{center}
\begin{tabular}{|c|c|c|}
\hline
Approaches&fine-tuning/prompting&EA\\
\hline
RESDSQL&fine-tuning&0.841\\
\hline
Graphix-T5&fine-tuning&0.793\\
\hline
T5-base + Tok&fine-tuning&0.756\\
\hline
RATSQL+GAP+NatSQL&fine-tuning&0.750\\
\hline
Dail-SQL&fine-tuning +fewshot&0.824\\
\hline
DIN-SQL&fewshot&0.742\\
\hline
C3&zeroshot&81.8\\
\hline
ChatGPT&zeroshot&0.767\\
\hline
SQL0+GPT4&zeroshot&0.763\\
\hline
SQL0+GPT4-turbo&zeroshot&0.768\\
\hline
SQL1+GPT4&zeroshot&0.792\\
\hline
SQL1+GPT4-turbo&zeroshot&0.797\\
\hline
SQL2+GPT4&zeroshot&0.811\\
\hline
SQL2+GPT4-turbo&zeroshot&0.812\\
\hline
SCVSQL+GPT4&zeroshot&0.821\\
\hline
SCVSQL+GPT4-turbo&zeroshot&0.824\\
\hline
\end{tabular}
\end{center}
\end{table}

We observed that when using only the Initial SQL Generation module, our method outperformed many mainstream methods, validating the quality of our prompts. Additionally, when employing the SQL-to-Schema method twice, namely, SCVSQL+GPT4-turbo, our approach surpassed all current zero shot and few shots prompting methods. The Dail-SQL method undergoes model training to select shot examples, and its few-shot setting consumes a substantial number of expensive GPT-4 tokens. In contrast, our method achieves comparable performance with minimal token consumption, showcasing the exceptional efficiency of our Text-to-SQL approach. While our method is only 0.019 lower than the optimal fine-tuning method, the RESDSQL method requires days of training time and significant GPU resources, whereas our approach only needs a multi-threaded CPU, completing all 1034 questions in the Spider Dev dataset within 10 minutes.

By analyzing Table 3, we find that the SQL-to-Schema strategy consistently contributes gains to SQL generation. Observing the frameworks of all zero shot and few shots methods, we can conclude that when employing prompting methods, a method incorporating numerous complex algorithmic modules may degrade the performance of large models. Conversely, using simpler, lightweight algorithmic modules can lead to remarkable performance on large models.

In order to further explore the upper limits of language models, for a specific problem, using our SQL0, SQL1, SQL2, SCVSQL methods under the same language model, we conducted experiments. If a language model can correctly answer a specific question with the appropriate strategy and prompt, we believe that it has the capability to solve that question. We recorded the maximum detection count of correct answers for the three models. The statistical results are presented in Table 4.

\begin{table}[htbp]
\caption{The maximum detection count of correct answers for different language models.}
\begin{center}
\begin{tabular}{|c|c|}
\hline
Large language model&upper limit\\
\hline
codellama-34B&0.8288\\
\hline
GPT4&0.8559\\
\hline
GPT4-turbo&0.8665\\
\hline
\end{tabular}
\label{tab1}
\end{center}
\end{table}
Although the accuracy of our four SQL generation methods gradually improves, However, the set of correctly answered questions by the less effective method is not a strict subset of the set of correctly answered questions by the more effective method. This tells us that language models heavily depend on prompts, and all current prompting methods have not yet reached the upper limit of the language model's capabilities. There is still significant room for improvement in the use of prompting methods in the Text-to-SQL domain.

\section{Conclusion}

We introduced the SQL-to-Schema method for the first time by defining table-recall@4 and demonstrated the efficiency of this schema linking method on large Language models. Combined with the SQL generation module, we outperformed all prompting methods in Text-to-SQL tasks. This suggests that the approach of extracting schema using initial SQL can bring global benefits to the Text-to-SQL task, extending beyond the schema linking phase. It further confirms the necessity of schema linking methods in Text-to-SQL tasks. Additionally, our SQL generation performance indicates that when leveraging large language models, there is no need for complex modules typically used in traditional machine learning fine-tuning methods. Simple strategies suffice for achieving excellent results. The full potential of large Language models has not yet been fully explored, and there is still room for exploration in schema linking or Text-to-SQL tasks. In future work, utilizing results generated with different strategies and prompts for voting may be a direction for our research.

\end{document}